\documentclass[superscriptaddress, prd, aps,amsmath,amssymb,showpacs,showkeys, twocolumn]{revtex4-1}
\usepackage[dvips]{graphicx,color}
\usepackage{subfig}
\usepackage{times}
\usepackage{xcolor}
\usepackage[%
  colorlinks=true,
  urlcolor=blue,
  linkcolor=red,
  citecolor=blue
]{hyperref}
\usepackage{orcidlink}

\begin{document}
\title{Equatorial periodic orbits and gravitational wave signatures in Euler–Heisenberg black holes surrounded by perfect fluid dark matter}

\author{Dhruba Jyoti Gogoi\orcidlink{0000-0002-4776-8506}}
\email[Email: ]{moloydhruba@yahoo.in}
\affiliation{Department of Physics, Madhabdev University, Narayanpur, Lakhimpur 784164, Assam, India}
\affiliation{Research Center of Astrophysics and Cosmology, Khazar University, Baku, AZ1096, 41 Mehseti Street, Azerbaijan}

\author{Jyatsnasree Bora\orcidlink{0000-0001-9751-5614}}
\email[Email: ]{jyatnasree.borah@gmail.com}
\affiliation{Department of Physics, Madhabdev University, Narayanpur, Lakhimpur 784164, Assam, India}
\affiliation{Pacif Institute of Cosmology and Selfology (PICS) Sagara, Sambalpur 768224, Odisha, India}

\author{Ali \"Ovg\"un \orcidlink{0000-0002-9889-342X}}
\email[Email: ]{ali.ovgun@emu.edu.tr}
\affiliation{Physics Department, Faculty of Arts and Sciences, Eastern Mediterranean University, Famagusta, 99628 North Cyprus via Mersin 10, Turkiye.}

%\date{}
\begin{abstract}
We investigate equatorial periodic orbits and their gravitational wave radiation in the spacetime of an Euler--Heisenberg (EH) black hole surrounded by perfect fluid dark matter (PFDM). The combined effects of quantum electrodynamic corrections and dark matter are incorporated through an effective metric, and the dynamics of timelike geodesics are analyzed using the effective potential formalism. We derive the conditions for marginally bound and innermost stable circular orbits, classify periodic trajectories using the rational parameter and topological indices, and identify a rich hierarchy of zoom--whirl motions in the strong-field regime. Gravitational wave signals from periodic orbits are computed using the numerical kludge method, revealing characteristic burst-like features associated with whirl phases. Our results show that perfect fluid dark matter systematically modifies the stability thresholds and suppresses the waveform amplitude, while QED corrections enhance high-frequency components generated near the horizon. These findings demonstrate that periodic orbits in the EH--PFDM spacetime provide a sensitive probe of quantum corrections and dark matter effects in strong gravitational fields.
\end{abstract}

%\begin{abstract}
%Work is based on Ref. \cite{Rani:2024qju, Rodrigues:2022zph}.
%\end{abstract}

%\pacs{04.30.Tv, 04.50.Kd}

\maketitle
%\tableofcontents %commented due to issue with compiling

\section{Introduction}\label{sec1}

Black holes provide a unique arena for testing gravity in its most nonlinear regime. 
The recent development of horizon-scale imaging and gravitational wave astronomy has made it possible to confront black hole models through several complementary observables, including light deflection, black hole shadows, accretion signatures, orbital dynamics, and ringdown spectra. In particular, null geodesic observables such as weak and strong lensing, photon capture, and shadow formation have proved to be powerful probes of the near-horizon geometry and of possible departures from the standard Schwarzschild or Kerr descriptions \cite{Kudo:2024aak,Gao:2024ejs,Igata:2025plb,Kumar:2019ohr,Yang:2022btw,Wang:2023vcv}. At the same time, timelike geodesics of massive particles provide an equally important and complementary channel, since they encode the structure of the effective potential, the stability of bound motion, and the dynamical behavior of compact objects moving in strong gravitational fields.

A major theme in recent years has been the investigation of black holes embedded in nonvacuum environments, especially dark-matter halos, as well as black holes modified by nonlinear electrodynamics, effective quantum corrections, or alternative theories of gravity. On the dark-matter side, black hole geometries in quantum wave dark matter, generalized-uncertainty-principle-inspired dark-matter backgrounds, and Dekel--Zhao halo profiles have been studied using shadow observables, weak deflection angles, and Event Horizon Telescope (EHT) constraints \cite{Pantig:2022sjb,Pantig:2022toh,Ovgun:2023wmc,Ovgun:2025bol}. Rotating black hole models surrounded by dark-matter halos have also been analyzed through EHT observations and accretion-disk properties \cite{Uktamov:2025lsq}, while the ringdown of black holes embedded in a Burkert dark-matter halo has recently been examined as a probe of environmental effects in the gravitational wave channel \cite{Yang:2025pmv}. Related studies have shown that periodic-orbit waveforms in Schwarzschild black holes surrounded by a Dehnen-type dark-matter halo can carry clear signatures of the surrounding matter distribution \cite{Alloqulov:2025ucf}.

Strong-field signatures have also been widely explored in modified or nonstandard black hole spacetimes. These include shadow and quasinormal-mode analyses of rotating Einstein--Euler--Heisenberg black holes \cite{Lambiase:2024lvo}, quasiperiodic oscillations, weak-field lensing, and shadow observables in Symmergent gravity \cite{Rayimbaev:2022hca}, null geodesics, causal structure, and accretion in Lorentzian--Euclidean black holes \cite{Capozziello:2025wwl}, and the dynamical and shadow properties of quantum-corrected Schwarzschild black holes in effective field theories of gravity \cite{Wang:2025fmz}. The orbital dynamics of massive particles and epicyclic motion have likewise been studied in several backgrounds, including Euler--Heisenberg black holes in cold dark-matter halos, magnetically charged black holes, quantum-corrected black holes, and conformally coupled charged black holes \cite{Mustafa:2025lix,Mustafa:2024kjy,Mustafa:2024kly,Mustafa:2024yhv}. Periodic equatorial motion and waveform signatures have also been investigated in regular black holes and black-bounce geometries, further demonstrating the sensitivity of strong-field orbital observables to near-horizon modifications of the metric \cite{Alloqulov:2025bxh,Bragado:2025jrg}. In parallel, the Gauss--Bonnet method has remained an important geometrical tool for extracting lensing observables, and has recently been extended to spinning massive particles in generic static and spherically symmetric spacetimes \cite{Pantig:2026qcf}.

From a theoretical perspective, one well-motivated source of short-distance corrections to the classical charged black hole geometry is Euler--Heisenberg (EH) nonlinear electrodynamics. The EH effective action arises from one-loop quantum electrodynamics and captures vacuum-polarization corrections to Maxwell theory through higher-order electromagnetic invariants \cite{HeisenbergEuler:1936,Schwinger:1951}. When coupled to gravity, these corrections deform the Reissner--Nordstr\"om geometry in the strong-field region and modify the near-horizon structure of charged black holes \cite{MagosBreton:2020}. Since periodic motion and zoom--whirl behavior are especially sensitive to the detailed shape of the effective potential near unstable circular trajectories, EH corrections provide a natural setting in which quantum-induced modifications may become dynamically relevant.

The environmental sector is equally important. Although the microscopic nature of dark matter is still unknown, phenomenological models that incorporate its gravitational effect on compact objects provide a useful framework for exploring possible observational imprints. A particularly convenient description is perfect fluid dark matter (PFDM), which leads to a static and spherically symmetric deformation of the metric function and has been used extensively in studies of optical properties and orbital motion around black holes \cite{HouXuWang:2018}. The combination of EH nonlinear electrodynamics with PFDM therefore offers a concrete laboratory in which short-distance nonlinear electromagnetic effects and large-scale environmental matter can be examined simultaneously \cite{MaWangDengHu:2024}. More broadly, the idea that external matter distributions can reshape black hole geometry and thermodynamics has a long history. Classical examples include black holes surrounded by clouds of strings \cite{Letelier:1979ej}, while regular black holes such as the Bardeen spacetime admit a nonlinear-electrodynamic interpretation \cite{Ayon-Beato:2000mjt}. These directions have been combined in recent studies of Bardeen-like black holes in string-cloud backgrounds and of their thermodynamic behavior \cite{Rodrigues:2022zph,Rani:2024qju}, reinforcing the broader motivation for studying systems in which nonlinear electrodynamics and an external matter environment act together.

Among all strong-field probes, periodic timelike geodesics are especially informative. Bound motion in a relativistic black hole potential can be organized in terms of periodic orbits whenever the ratio of the azimuthal and radial frequencies is rational. These trajectories provide a natural skeleton for the classification of generic bound motion and reveal the transition from ordinary relativistic precession to the highly nontrivial zoom--whirl regime \cite{Grossman:2008yk,Levin:2008mq,Glampedakis:2002ya,Misra:2010pu,Levin:2009sk}. The zoom--whirl taxonomy, commonly described by the integers $(z,w,v)$, has been used extensively to map strong-field orbital structure in a wide range of black hole spacetimes and alternative gravity scenarios \cite{Levin:2008ci,Fujita:2009bp,Healy:2009zm,Wei:2019zdf,Azreg-Ainou:2020bfl,Deng:2020yfm,Deng:2020hxw,Wang:2022tfo,QiQi:2024dwc,Alloqulov:2025ucf,Haroon:2025rzx,Wang:2025hla,Lu:2025cxx,Tu:2023xab,Yang:2024lmj,Jiang:2024cpe,Huang:2025vpi}. In these systems, the orbital morphology is strongly tied to the effective potential and therefore to the underlying spacetime geometry.

This is particularly relevant for extreme mass-ratio inspirals (EMRIs), in which a stellar-mass compact object gradually inspirals into a supermassive black hole \cite{Xamidov:2026kqs,Burko:2006ua}. In the adiabatic regime, the motion may be viewed as a sequence of slowly evolving bound geodesics, and near-critical portions of the orbit are often well captured by periodic or zoom--whirl trajectories. Since the emitted gravitational radiation depends sensitively on the orbital frequencies, periapsis structure, and near-horizon motion, modifications of the background geometry can leave distinct imprints on the waveform \cite{Hughes:2000ssa,Babak:2017tow}. This makes periodic orbits an especially useful theoretical bridge between geodesic dynamics and observable gravitational wave features.

Motivated by these developments, the purpose of the present work is to study equatorial periodic orbits and their gravitational wave signatures in the spacetime of an Euler--Heisenberg black hole surrounded by perfect fluid dark matter. Our focus is on the combined effect of nonlinear electrodynamic corrections and a dark-matter environment on the structure of timelike bound motion in the strong-field region. In particular, the PFDM sector modifies the geometry on intermediate and larger scales and shifts the characteristic orbital radii, whereas the EH correction becomes important at smaller radii and can alter the near-horizon part of the motion where whirl episodes and sharp waveform bursts are generated. Although shadows, lensing, QPOs, epicyclic oscillations, accretion properties, and quasinormal modes have been investigated in a variety of related black hole backgrounds, a detailed study of periodic timelike motion and its associated waveform morphology in the EH--PFDM spacetime is still lacking.

In this paper, we first analyze equatorial timelike geodesics using the effective-potential formalism and determine the marginally bound orbit and the innermost stable circular orbit (ISCO). We then classify periodic bound trajectories through the rational parameter and the topological indices associated with zoom--whirl motion. Finally, we construct representative gravitational wave signals from these periodic orbits using the numerical kludge approach, with the aim of isolating how the PFDM parameter and the EH nonlinear-electrodynamic correction affect orbital morphology and waveform structure. In this way, the present work connects strong-field geodesic dynamics to gravitational wave phenomenology in a black hole spacetime that incorporates both quantum-inspired electromagnetic corrections and an external dark-matter environment.

This paper is organized as follows. In Sec.~\ref{sec02}, we summarize the EH--PFDM spacetime and its main geometric properties. In Sec.~\ref{sec03}, we investigate equatorial timelike motion, determine the relevant orbital thresholds, and classify the periodic bound trajectories. In Sec.~\ref{sec04}, we compute numerical-kludge gravitational waveforms for representative periodic orbits and analyze their dependence on the model parameters. We conclude in Sec.~\ref{sec05}. Throughout the paper, we use units $G=c=\hbar=1$.

\section{EH Black Hole Surrounded by Perfect Fluid Dark Matter}\label{sec02}
Following the introduction section, in this section, we assemble the spacetime geometry describing an EH black hole dipped in perfect fluid dark matter. The solution is obtained by combining the effects of non-linear electrodynamics arising from quantum electrodynamic corrections with the gravitational contribution of dark matter, treated as an independent perfect fluid source. The solution was introduced in Ref. \cite{Ma:2024oqe}.

\subsection{Action and Field Equations}
The total action for black holes with EH–PFDM spacetime can be given as \cite{Ma:2024oqe}
\begin{equation}
S=\int d^4x \sqrt{-g}
\left[
\frac{R}{16\pi}
+\mathcal{L}_{\rm EH}(F)
+\mathcal{L}_{\rm PFDM}
\right],
\label{action}
\end{equation}
where $R$ is the Ricci scalar, $\mathcal{L}_{\rm EH}$ denotes the EH nonlinear electromagnetic Lagrangian, and $\mathcal{L}_{\rm PFDM}$ represents the effective Lagrangian of perfect fluid dark matter. The electromagnetic invariant is defined as $F=F_{\mu\nu}F^{\mu\nu}$.

The EH Lagrangian up to leading-order QED correction takes the form \cite{EHtheory,EHsolution}
\begin{equation}
\mathcal{L}_{\rm EH}(F)
=-\frac{1}{4}F
+\frac{a}{8}F^2,
\label{EHL}
\end{equation}
where the parameter $a$ encodes the strength of quantum electrodynamic corrections.

Varying the action \eqref{action} with respect to the metric yields the Einstein field equations,
\begin{equation}
G_{\mu\nu}=8\pi
\left(
T_{\mu\nu}^{\rm EH}
+T_{\mu\nu}^{\rm PFDM}
\right),
\label{EFE}
\end{equation}
where the total energy-momentum tensor is the sum of contributions from nonlinear electrodynamics and dark matter.

\subsection{Metric Ansatz and Solution}

We assume a spacetime which is static, spherically symmetric in nature. 
\begin{equation}
ds^2
=-g(r)\,dt^2
+\frac{dr^2}{g(r)}
+r^2(d\theta^2+\sin^2\theta\,d\phi^2).
\label{metric}
\end{equation}

For the EH electromagnetic field, the energy-momentum tensor in the purely electric case is given by \cite{EHbH1}
\begin{equation}
T^{t}_{\ t\,({\rm EH})}
=\frac{1}{4\pi}
\left(
-\frac{Q^2}{2r^4}
+\frac{aQ^4}{8r^8}
\right),
\label{TEH}
\end{equation}
where $Q$ is the electric charge of the black hole.

For perfect fluid dark matter, the effective energy-momentum tensor reads \cite{DMVI}
\begin{equation}
T^{t}_{\ t\,({\rm PFDM})}
=\frac{\alpha}{8\pi r^3},
\label{TPFDM}
\end{equation}
where $\alpha$ is the dark matter parameter characterizing the density profile.

Substituting Eqs.~\eqref{TEH} and \eqref{TPFDM} into the Einstein equations \eqref{EFE}, the metric function $g(r)$ is obtained as \cite{Ma:2024oqe}
\begin{equation}
g(r)=
1-\frac{2M}{r}
+\frac{Q^2}{r^2}
-\frac{aQ^4}{20r^6}
+\frac{\alpha}{r}\ln\!\left(\frac{r}{|\alpha|}\right),
\label{grfinal}
\end{equation}
where $M$ is the ADM mass of the black hole.

The first three terms correspond to the standard Reissner–Nordstr\"om geometry, the fourth term arises from the leading-order QED correction, and the logarithmic term encodes the contribution of perfect fluid dark matter.

\subsection{Limiting Cases and Physical Interpretation}

Several physically relevant limits can be recovered from Eq.~\eqref{grfinal}:

\begin{itemize}
\item For $a\to0$ and $\alpha\to0$, the metric reduces to the Reissner–Nordstr\"om solution.
\item For $Q\to0$, one obtains a Schwarzschild black hole surrounded by PFDM.
\item For $\alpha\to0$, the solution reduces to the standard Euler-Heisenberg black hole.
\end{itemize}

The logarithmic PFDM term dominates at large distances and modifies the asymptotic structure of the spacetime, while the QED correction term becomes significant only at small radii. Therefore, the EH parameter primarily affects near-horizon physics, whereas dark matter influences the global geometry and particle motion at intermediate and large scales.

\section{Particle Orbits in EH Black Hole with PFDM}\label{sec03}

In this section, we investigate the motion of test particles in the background of the EH black hole surrounded by perfect fluid dark matter, described by the metric \eqref{grfinal}. The dynamics of particle motion is governed by the geodesic equations, which can be derived from the Lagrangian of a free particle in curved spacetime,
\begin{equation}
\mathcal{L}=\frac{1}{2} g_{\mu\nu}\dot{x}^{\mu}\dot{x}^{\nu}=\delta,
\label{lagrangian_EH}
\end{equation}
where the overdot denotes differentiation with respect to the affine parameter $\tau$, and $\delta=-1$ and $0$ correspond to massive and massless particles, respectively. Due to spherical symmetry, we restrict motion to the equatorial plane by setting $\theta=\pi/2$ and $\dot{\theta}=0$.

The generalized four-momentum associated with the Lagrangian \eqref{lagrangian_EH} is
\begin{equation}
p_{\mu}=\frac{\partial \mathcal{L}}{\partial \dot{x}^{\mu}}=g_{\mu\nu}\dot{x}^{\nu}.
\end{equation}
Since the metric is independent of $t$ and $\phi$, the corresponding momenta are conserved,
\begin{gather}
p_t=g_{tt}\dot{t}=-E, \\
p_\phi=g_{\phi\phi}\dot{\phi}=L, \\
p_r=g_{rr}\dot{r}, \\
p_\theta=0,
\end{gather}
where $E$ and $L$ represent the conserved energy and angular momentum per unit mass.

Using the metric \eqref{grfinal}, the first integrals of motion become
\begin{gather}
\dot{t}=\frac{E}{g(r)}, \\
\dot{\phi}=\frac{L}{r^2}, \\
\dot{r}^2=g(r)\left[\frac{E^2}{g(r)}-\left(\delta+\frac{L^2}{r^2}\right)\right].
\label{rdot_EH}
\end{gather}

For timelike geodesics ($\delta=-1$), the radial equation can be written in the effective potential form,
\begin{equation}
\dot{r}^2+V_{\mathrm{eff}}(r)=E^2,
\label{rdoteff_EH}
\end{equation}
with the effective potential
\begin{equation}
V_{\mathrm{eff}}(r)=g(r)\left(1+\frac{L^2}{r^2}\right).
\label{veff_EH}
\end{equation}

The effective potential encodes the combined influence of QED correction and dark matter on particle motion. The allowed regions of motion are determined by $E^2\geq V_{\mathrm{eff}}(r)$, while circular and bound orbits correspond to extrema of $V_{\mathrm{eff}}(r)$.

\begin{figure*}[!ht]
\centering
\includegraphics[scale=0.55]{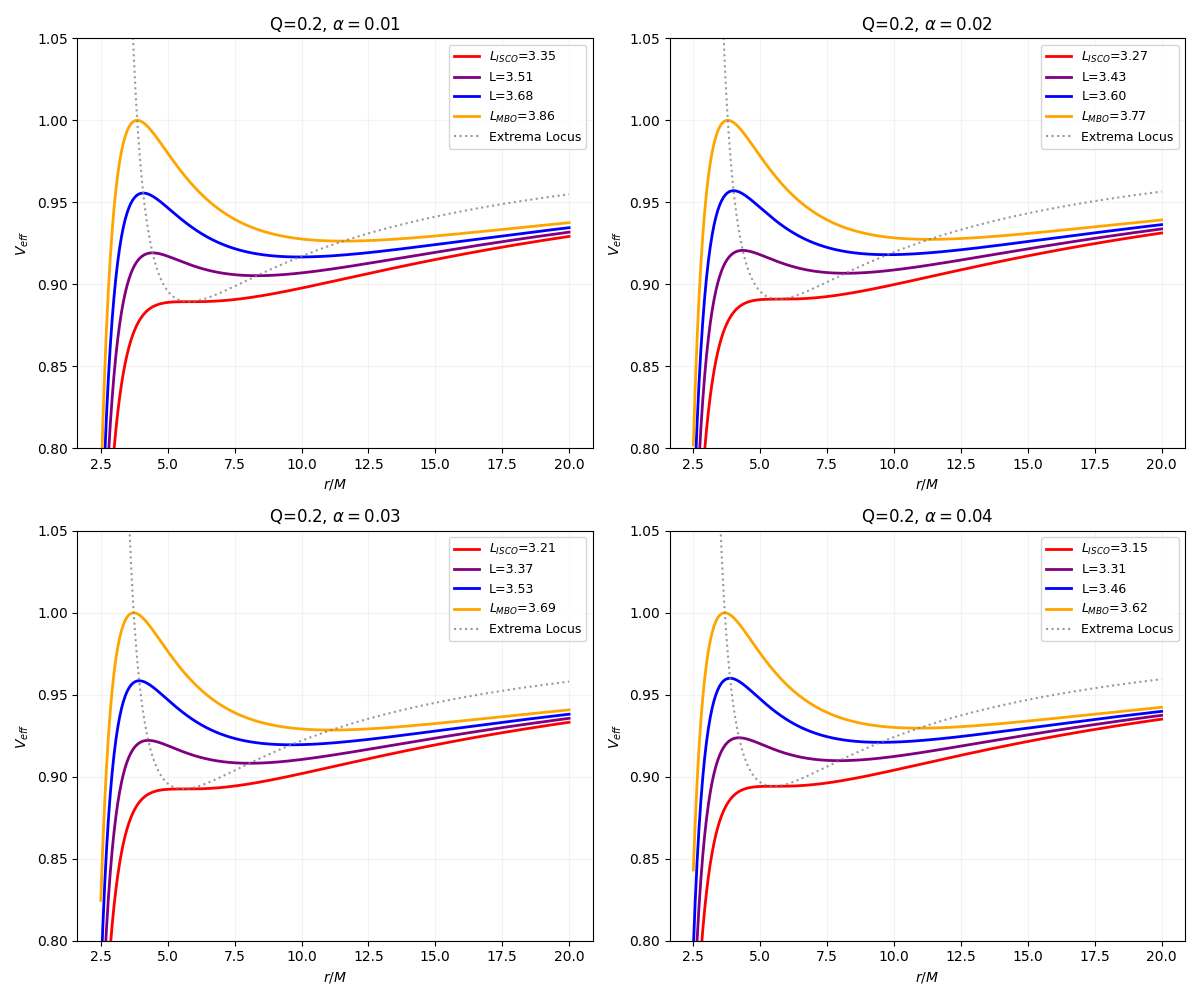}
\caption{Effective potential $V_{\mathrm{eff}}(r)$ for the EH-PFDM black hole for different values of the model parameters. The curves correspond to representative values of the angular momentum, including $L_{\mathrm{ISCO}}$ and $L_{\mathrm{MBO}}$. The dotted line denotes the locus of extrema of the potential.}
\label{fig:potential}
\end{figure*}

Figure~\ref{fig:potential} illustrates the effective potential  $V_{\mathrm{eff}}(r)$ for timelike geodesics in the EH-PFDM spacetime for  several values of the dark matter parameter $\alpha$, with the electric charge fixed at $Q=0.2$. Each panel shows representative curves corresponding to different values of the angular momentum, including the innermost stable circular orbit $L_{\mathrm{ISCO}}$ and the marginally bound orbit $L_{\mathrm{MBO}}$. The dotted line denotes the locus of extrema of the potential, which separates stable and unstable circular trajectories.

For a given $\alpha$, the effective potential exhibits a local minimum and a local maximum. The minimum corresponds to a stable circular orbit, while the maximum represents an unstable circular orbit associated with the photon sphere. Bound motion is allowed in the region where $E^2$ lies between these two extrema, whereas plunging or unbound trajectories occur when the energy 
exceeds the potential barrier.

As the dark matter parameter $\alpha$ increases, both the height of the 
potential barrier and the depth of the potential well decrease systematically. 
Physically, this indicates that perfect fluid dark matter weakens the effective 
gravitational attraction experienced by the particle, leading to less tightly 
bound orbits. Consequently, the locations of both the ISCO and MBO shift 
outward, and the corresponding angular momenta $L_{\mathrm{ISCO}}$ and 
$L_{\mathrm{MBO}}$ decrease monotonically with increasing $\alpha$.

The dotted extrema locus reveals that the radius at which the potential 
extremum occurs increases with $\alpha$, confirming that dark matter 
effectively pushes the strong-field region to larger distances from the 
black hole. This behaviour implies that accretion disks and periodic orbits in 
EH-PFDM spacetimes are expected to be more extended than in the standard 
Reissner-Nordstr\"Om or Schwarzschild geometries.

Overall, Fig.~\ref{fig:potential} demonstrates that perfect fluid dark matter 
introduces a controlled deformation of the effective potential, reducing the 
stability domain of bound orbits and modifying the characteristic radii of 
strong-field motion. These effects directly influence the orbital structure, 
accretion dynamics, and gravitational wave emission from compact objects 
around EH-PFDM black holes.

\subsection{Marginally Bound Orbits}\label{s4sub1}

A marginally bound orbit (MBO) is defined by the limiting case of unit energy, $E=1$. The conditions for such an orbit are
\begin{equation}
V_{\mathrm{eff}}(r)=1,\qquad 
\frac{dV_{\mathrm{eff}}}{dr}=0.
\label{mbo_EH}
\end{equation}
Solving these equations simultaneously yields the radius $r_{\mathrm{MBO}}$ and angular momentum $L_{\mathrm{MBO}}$. Both quantities depend explicitly on the QED parameter $a$ and the dark matter parameter $\alpha$, implying that quantum electrodynamic corrections and PFDM jointly modify the location of the marginally bound orbit compared with the RN and Schwarzschild cases.

\subsection{Innermost Stable Circular Orbits}\label{s4sub2}

The innermost stable circular orbit (ISCO) represents the smallest radius for which stable circular motion is possible. It is determined by the conditions
\begin{equation}
V_{\mathrm{eff}}(r)=E^2,\qquad 
\frac{dV_{\mathrm{eff}}}{dr}=0,\qquad 
\frac{d^2V_{\mathrm{eff}}}{dr^2}=0.
\label{isco_EH}
\end{equation}
From these equations, the ISCO angular momentum and energy are obtained as
\begin{gather}
L_{\mathrm{ISCO}}^2=
\frac{r_{\mathrm{ISCO}}^3 g'(r_{\mathrm{ISCO}})}
{2g(r_{\mathrm{ISCO}})-r_{\mathrm{ISCO}} g'(r_{\mathrm{ISCO}})}, \\
E_{\mathrm{ISCO}}^2=
\frac{g(r_{\mathrm{ISCO}})^2}
{g(r_{\mathrm{ISCO}})-r_{\mathrm{ISCO}} g'(r_{\mathrm{ISCO}})},
\end{gather}
where the prime denotes differentiation with respect to $r$.

The ISCO radius satisfies
\begin{equation}
2g(r_{\mathrm{ISCO}})
-r_{\mathrm{ISCO}} g'(r_{\mathrm{ISCO}})
-r_{\mathrm{ISCO}}^2 g''(r_{\mathrm{ISCO}})=0.
\label{riscocond_EH}
\end{equation}

In the limit $\alpha\to0$ and $a\to0$, these expressions reduce to the standard Schwarzschild ISCO at $r_{\mathrm{ISCO}}=6M$. The presence of PFDM generally shifts the ISCO outward, while QED corrections introduce additional deviations at small radii.

\begin{figure}[!htbp]
\centering
\includegraphics[scale=0.55]{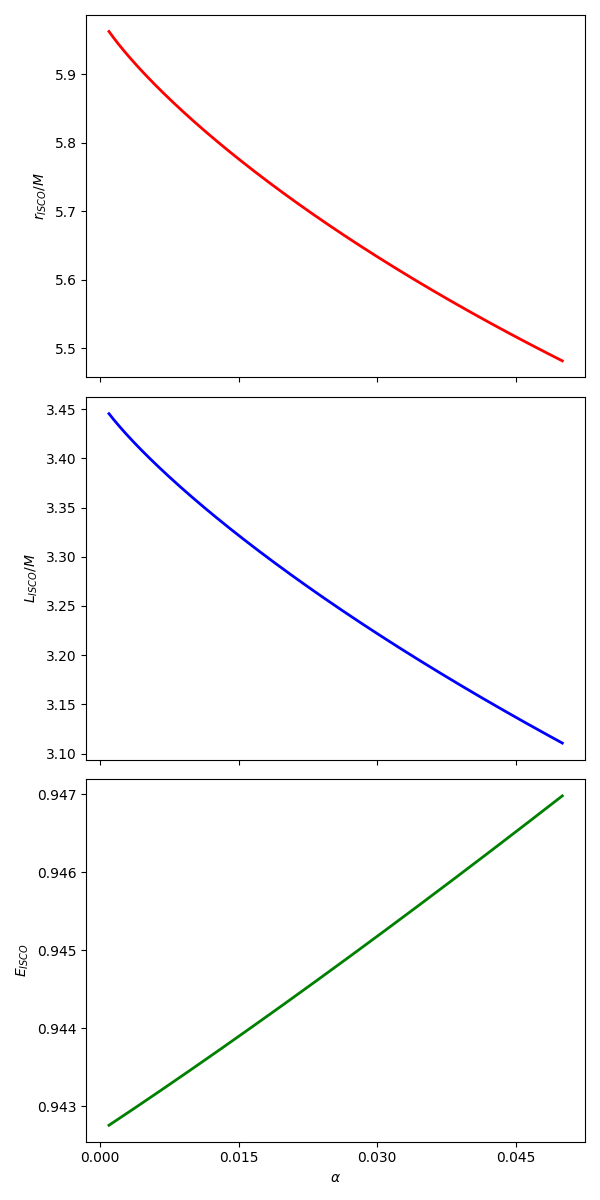}
\caption{Dependence of the ISCO radius $r_{\mathrm{ISCO}}$, ISCO angular momentum $L_{\mathrm{ISCO}}$, and ISCO energy $E_{\mathrm{ISCO}}$ on the model parameter $\alpha$.}
\label{fig:isco}
\end{figure}

\begin{figure}[!htbp]
\centering
\includegraphics[scale=0.5]{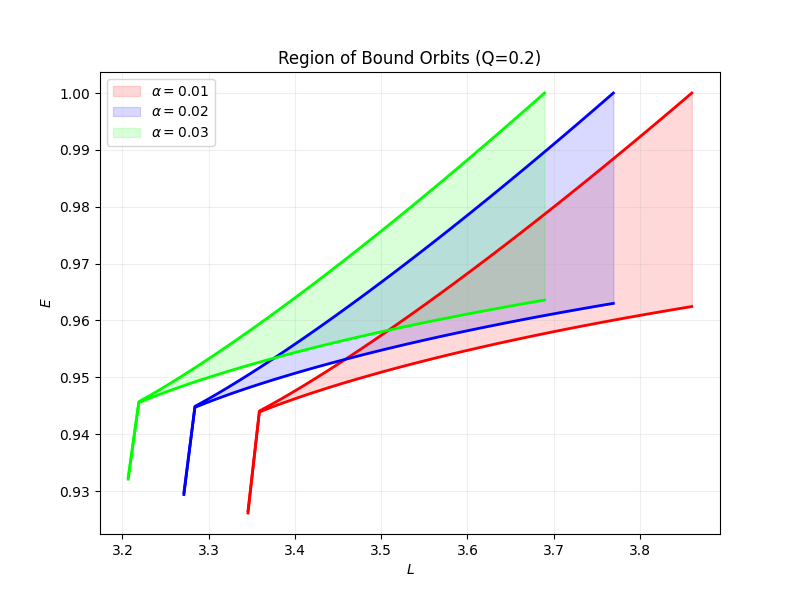}
\caption{Allowed parameter regions for bound timelike orbits in the $E$--$L$ plane for different values of the model parameter $\alpha$.}
\label{fig:param}
\end{figure}

Figure~\ref{fig:isco} displays the dependence of the ISCO radius $r_{\mathrm{ISCO}}$, 
the ISCO angular momentum $L_{\mathrm{ISCO}}$, and the ISCO energy $E_{\mathrm{ISCO}}$ 
on the perfect fluid dark matter parameter $\alpha$. As $\alpha$ increases, both 
$r_{\mathrm{ISCO}}$ and $L_{\mathrm{ISCO}}$ decrease monotonically, while 
$E_{\mathrm{ISCO}}$ increases. This behaviour indicates that the presence of 
perfect fluid dark matter weakens the effective gravitational attraction, allowing 
stable circular orbits to exist closer to the black hole with smaller angular 
momentum, but requiring slightly higher energy to maintain stability. Physically, 
this reflects a redistribution of the effective potential induced by dark matter, 
which flattens the potential well and shifts the stability threshold inward.

Figure~\ref{fig:param} shows the allowed parameter regions for bound timelike 
orbits in the $(E,L)$ plane for different values of $\alpha$. The shaded areas 
represent the domain in which bound motion is possible, bounded from below by 
the ISCO curve and from above by the marginally bound orbit. As $\alpha$ increases, 
the admissible region shrinks and shifts toward lower angular momentum values. 
This implies that dark matter reduces the phase space available for long-lived 
bound orbits, making strongly bound trajectories less likely and facilitating 
plunging or escape orbits for a wider range of initial conditions.

\subsection{Periodic Orbits and Rational Parameter}\label{s4sub3}

Periodic orbits correspond to bound trajectories for which the ratio of angular and radial frequencies is rational. The rational parameter $q$ is defined as
\begin{equation}
q=\frac{\omega_\phi}{\omega_r}-1
=\frac{\Delta\phi}{2\pi}-1,
\label{qdef_EH}
\end{equation}
where $\Delta\phi$ is the total azimuthal angle accumulated during one complete radial oscillation between the periapsis $r_p$ and apoapsis $r_a$,
\begin{equation}
\Delta\phi
=2\int_{r_p}^{r_a}
\frac{L}{r^2\sqrt{E^2-g(r)\left(1+\frac{L^2}{r^2}\right)}}\,dr.
\label{deltaphi_EH}
\end{equation}

The rational number $q$ can be written in terms of three integers $(z,w,v)$ \cite{Levin:2008mq},
\begin{equation}
q=w+\frac{v}{z},
\label{zwv_EH}
\end{equation}
where $z$ is the zoom number, $w$ is the number of whirls, and $v$ is the vertex number. This zoom-whirl structure characterizes the topology of strong-field orbits around the EH–PFDM black hole.

For bound motion, the angular momentum is restricted to
\begin{equation}
L_{\mathrm{ISCO}}\leq L\leq L_{\mathrm{MBO}},
\end{equation}
and it is convenient to parametrize it as
\begin{equation}
L=L_{\mathrm{ISCO}}
+\epsilon\left(L_{\mathrm{MBO}}-L_{\mathrm{ISCO}}\right),
\qquad 0<\epsilon<1.
\label{epsilon_EH}
\end{equation}
By fixing $\epsilon$ and solving Eq.~\eqref{qdef_EH} for the energy $E$, one can systematically construct families of periodic orbits and analyze how QED corrections and dark matter modify their geometric structure.

\begin{figure}[htbp]
\centering
\includegraphics[scale=0.45]{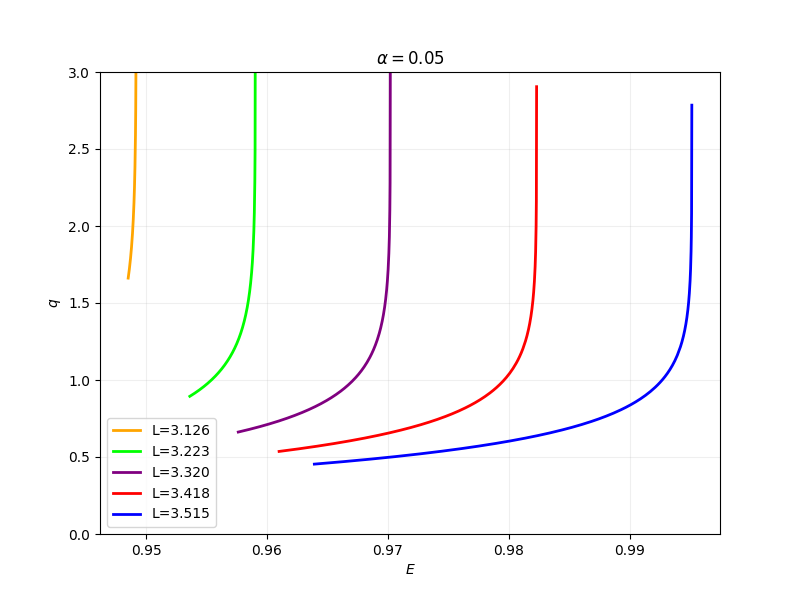}
\caption{Rational parameter $q$ as a function of the particle energy $E$ for fixed angular momentum in the EHPFDM black hole.}
\label{fig:qE}
\end{figure}

Figure~\ref{fig:qE} illustrates the rational parameter $q$ as a function of the 
particle energy $E$ for several fixed values of the angular momentum. The 
parameter $q$ diverges as $E$ approaches its critical value associated with the 
unstable circular orbit, signaling the onset of extreme zoom--whirl behaviour. 
For larger $L$, the divergence occurs at higher energies, indicating that 
stronger angular momentum delays the transition to whirl-dominated motion. 
This demonstrates that the topological structure of periodic orbits is highly 
sensitive to the energetic proximity to the unstable orbit, and that the 
EH--PFDM spacetime supports a rich hierarchy of zoom--whirl trajectories.

\begin{figure*}[!htbp]
\centering
\includegraphics[scale=0.45]{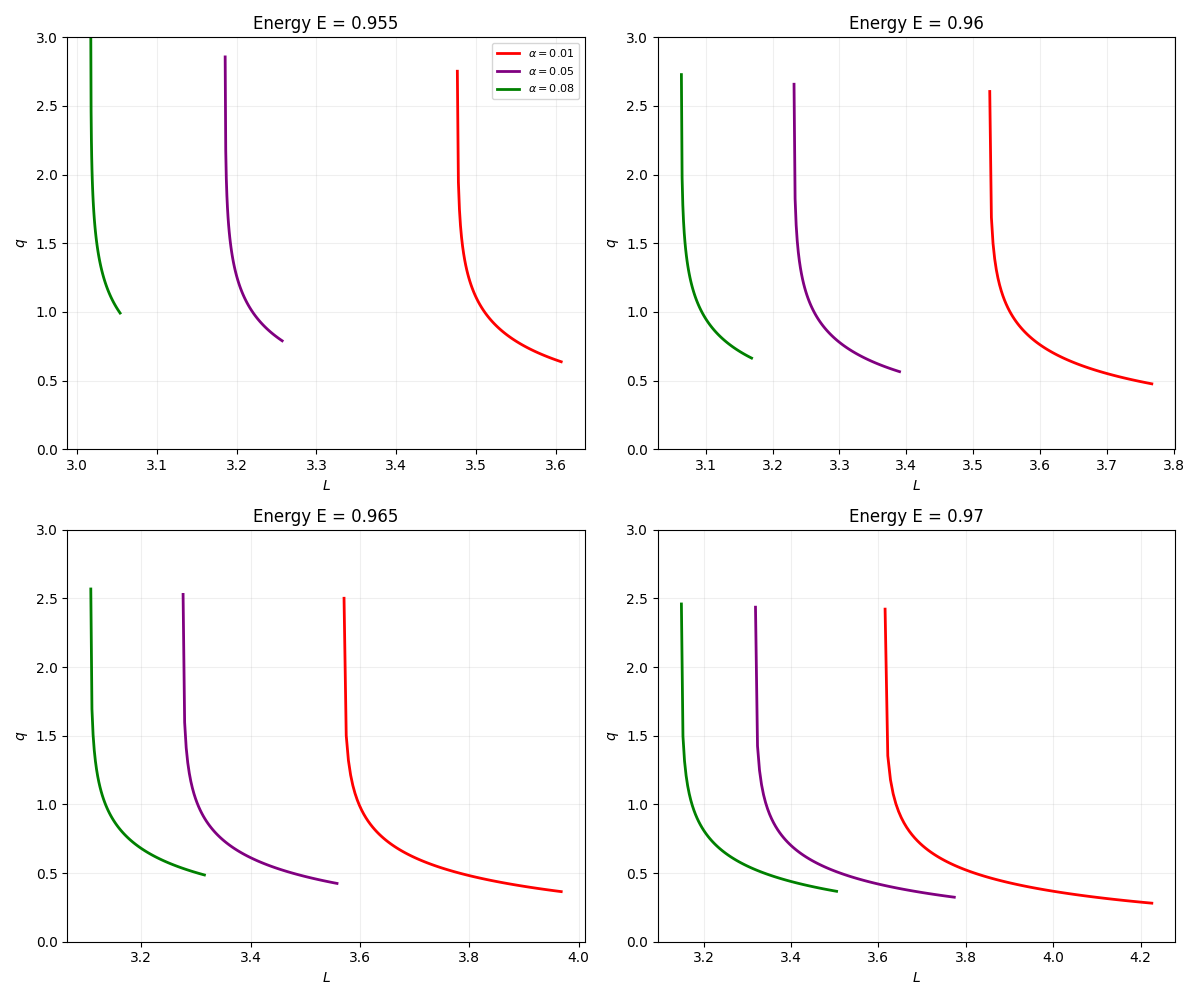}
\caption{Rational parameter $q$ as a function of the angular momentum $L$ for fixed energy in the EHPFDM black hole.}
\label{fig:qL}
\end{figure*}

Figure~\ref{fig:qL} shows the rational parameter $q$ as a function of the angular 
momentum $L$ for several fixed values of the particle energy $E$ in the EH--PFDM 
spacetime. Each panel corresponds to a different energy level, while the coloured 
curves represent different values of the perfect fluid dark matter parameter 
$\alpha$. The rational parameter $q$ encodes the ratio between the azimuthal and 
radial frequencies and thus provides a topological classification of periodic 
orbits.

For fixed energy, $q$ decreases monotonically as the angular momentum increases. 
This behaviour reflects the fact that larger angular momentum suppresses 
whirl motion and promotes more regular, nearly Keplerian trajectories. In the 
limit of large $L$, the orbits become weakly relativistic and the precession 
per radial cycle decreases, leading to smaller values of $q$.

In contrast, as $L$ approaches its critical lower bound associated with the 
unstable circular orbit, the rational parameter diverges, $q\rightarrow\infty$. 
This divergence signals the onset of extreme zoom--whirl behaviour, in which the 
particle executes an arbitrarily large number of revolutions near the unstable 
orbit before escaping back to the outer region. This regime corresponds to 
strong-field dynamics dominated by the near-horizon geometry.

The effect of perfect fluid dark matter is clearly visible in the horizontal 
shift of the curves. Increasing $\alpha$ systematically moves the divergence 
point toward smaller values of $L$, indicating that dark matter facilitates 
the emergence of zoom--whirl motion at lower angular momentum. Physically, this 
is a consequence of the effective weakening of the central gravitational field, 
which modifies the location of the unstable circular orbit and alters the 
threshold for whirl-dominated dynamics.

As the energy increases from $E=0.955$ to $E=0.97$, the divergence points shift 
to larger values of $L$ and the overall magnitude of $q$ decreases. This implies 
that higher-energy particles require larger angular momentum to sustain bound 
whirl motion and that extreme zoom--whirl behaviour becomes less pronounced at 
higher energies.

\begin{figure*}[htbp]
\centering
\includegraphics[scale=0.55]{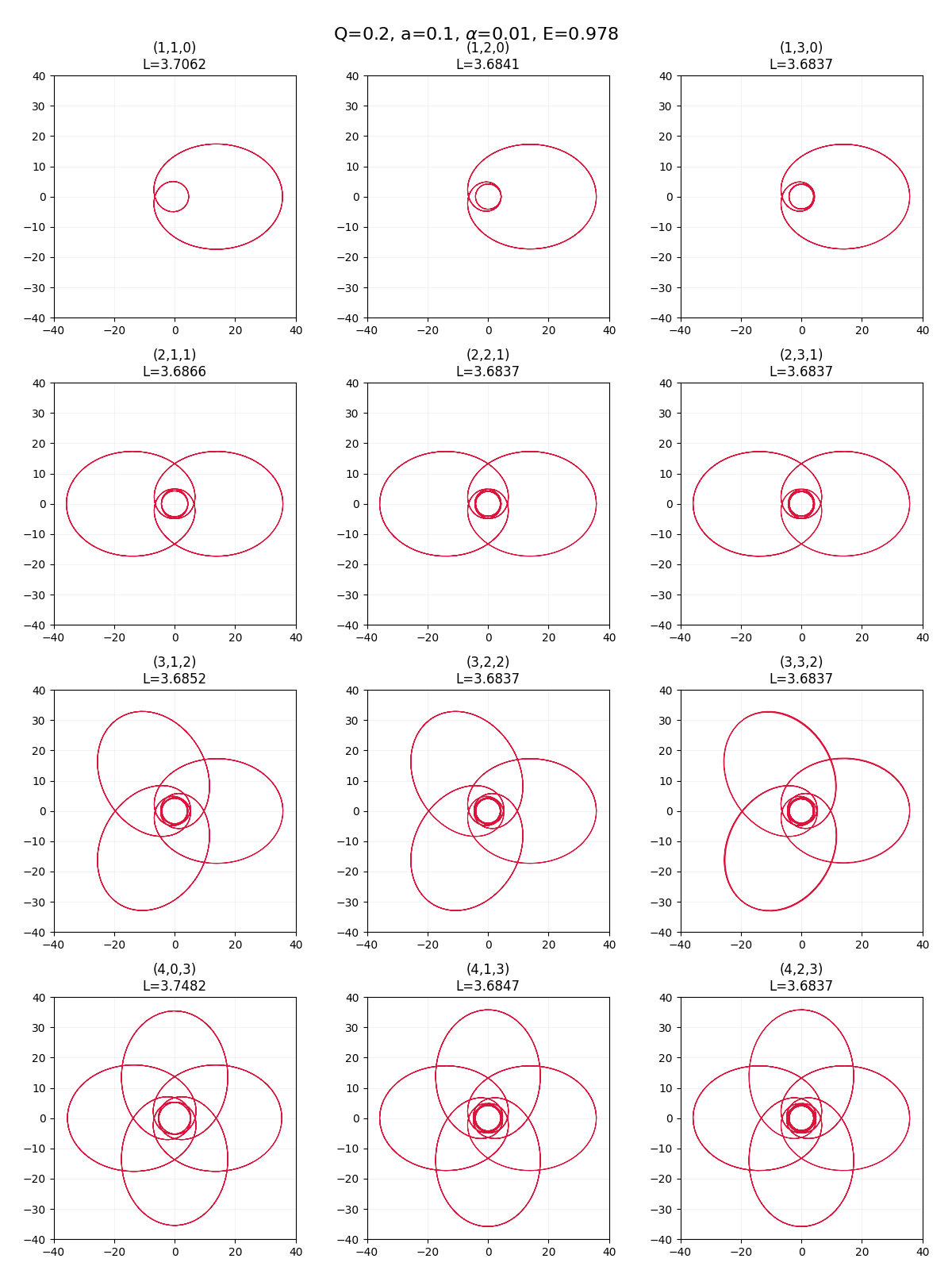}
\caption{Periodic orbits in the EHPFDM black hole for fixed energy, characterized by different values of $(z,w,\nu)$.}
\label{fig:wz1}
\end{figure*}

\begin{figure*}[htbp]
\centering
\includegraphics[scale=0.55]{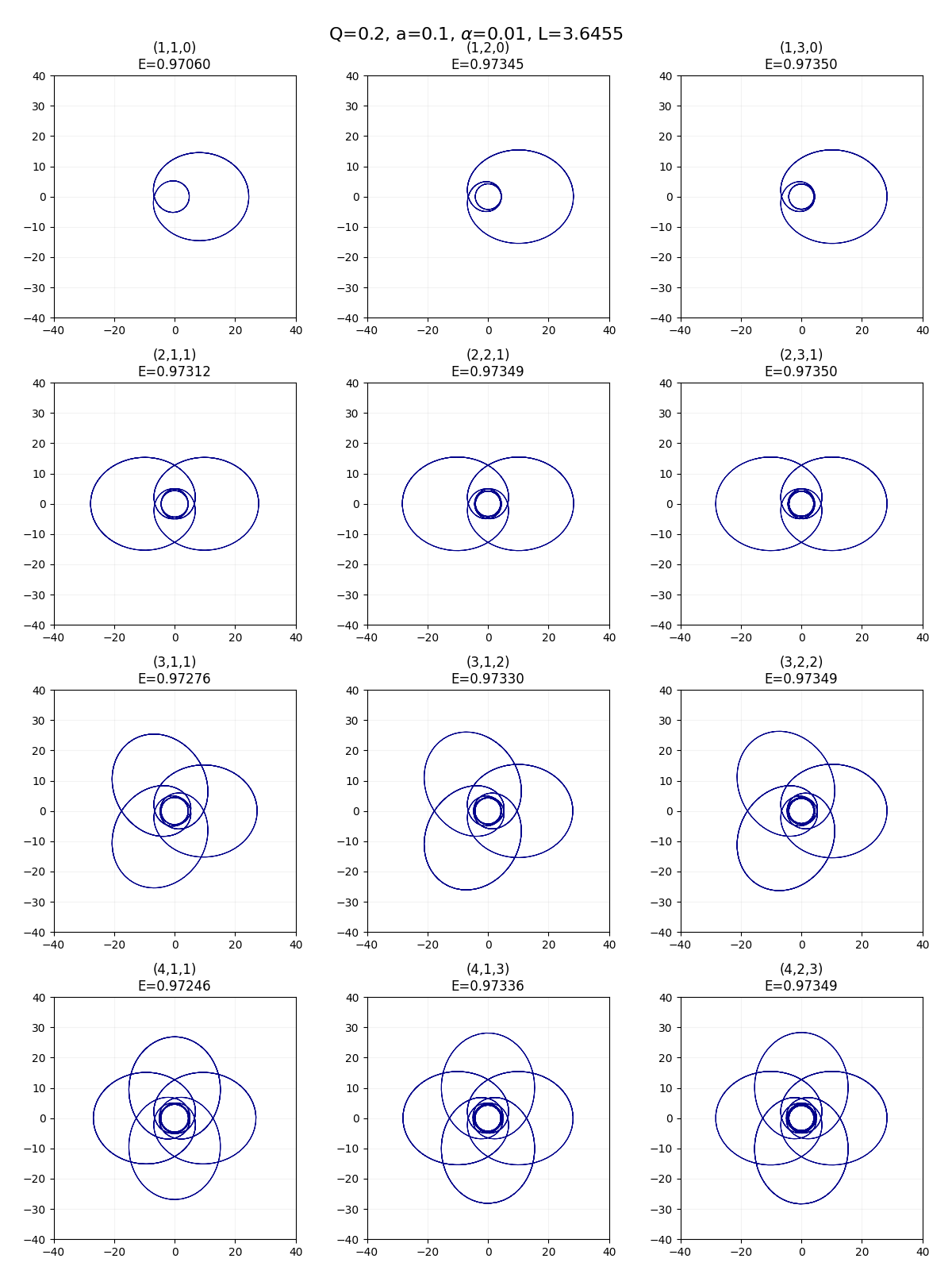}
\caption{Periodic orbits in the EHPFDM black hole for fixed angular momentum and different values of $(z,w,\nu)$.}
\label{fig:wz2}
\end{figure*}

Figures~\ref{fig:wz1} and \ref{fig:wz2} display representative families of 
periodic equatorial orbits in the EH--PFDM black hole, classified according 
to the topological indices $(z,w,\nu)$, which encode the zoom number, whirl 
number, and vertex number, respectively. These indices provide a complete 
topological characterization of relativistic periodic motion and allow 
a systematic organization of bound trajectories in the strong-field regime.

Figure~\ref{fig:wz1} corresponds to the case of fixed energy, where the angular 
momentum is tuned to realize different periodic orbits. As the zoom number 
$z$ increases, the outer envelope of the orbit develops additional lobes, 
reflecting the increasing number of radial oscillations before closure. 
Simultaneously, larger values of the whirl number $w$ produce more tightly 
wound loops near the center, indicating prolonged trapping of the particle 
close to the unstable circular orbit.

Physically, this behaviour reflects the hierarchical structure of the 
effective potential. Orbits with small $(z,w)$ are dominated by Newtonian-like 
motion, with only mild relativistic precession. In contrast, orbits with 
larger $w$ exhibit pronounced zoom--whirl dynamics, spending a significant 
fraction of their period in the strong-field region where relativistic 
effects dominate. The presence of multiple inner loops signals repeated 
temporary capture near the photon sphere, a purely general-relativistic 
phenomenon absent in weak-field gravity.

Figure~\ref{fig:wz2} shows periodic orbits for fixed angular momentum, where 
the energy is adjusted to realize different topological classes. In this case, 
increasing $(z,w,\nu)$ leads to progressively more intricate orbital patterns, 
with the particle exploring a larger portion of phase space while maintaining 
the same overall centrifugal support. Higher-energy orbits develop wider outer 
lobes and more extended trajectories, while still exhibiting strong whirl 
episodes near the black hole.

Comparing Figs.~\ref{fig:wz1} and \ref{fig:wz2}, one observes that varying the 
energy primarily controls the global size and eccentricity of the orbit, 
whereas varying the angular momentum mainly regulates the degree of whirl 
activity near the center. This demonstrates that energy governs the radial 
excursions of the motion, while angular momentum controls the depth of 
strong-field trapping.

\section{Numerical Kludge Gravitational Wave Radiation from Periodic Orbits}
\label{sec04}

In an extreme mass-ratio inspiral (EMRI) system, the Euler--Heisenberg black hole 
surrounded by perfect fluid dark matter, described by the metric function 
$g(r)$ in Eq.~\eqref{grfinal}, can effectively model a supermassive compact object. 
A small body moving along a periodic bound orbit in this strong-field spacetime 
acts as a source of gravitational radiation. Since the mass of the orbiting object 
is assumed to be much smaller than that of the central black hole, its backreaction 
on the geometry can be neglected and the motion may be treated as a perturbation 
on a fixed background.

In this regime, the adiabatic approximation is applicable 
\cite{Hughes:2000ssa, Glampedakis:2002ya, Zhang:2025wni}, 
which assumes that the loss of energy and angular momentum due to gravitational 
radiation is negligible over one orbital period. Consequently, the constants of 
motion $(E,L)$ may be treated as fixed while computing the waveform associated 
with a single periodic orbit. This approach has proven reliable for modelling 
EMRIs and has been widely adopted in recent studies 
\cite{Zhang:2025wni, Li:2025eln, Zhao:2024exh,Meng:2024cnq}.

We employ the numerical kludge method introduced in Ref.~\cite{BabakEtAl:2007}. 
The procedure consists of two steps: 
(i) the periodic timelike geodesics are obtained by numerically integrating the 
equations of motion in the EH--PFDM spacetime; 
(ii) the resulting trajectories are embedded into a pseudo-flat background and 
used as sources in the quadrupole formula to generate the gravitational waveform.

The small body oscillates between periapsis and apoapsis while executing repeated 
zoom--whirl motion near the unstable circular orbit. As the particle approaches 
the black hole, it undergoes multiple revolutions in a region of strong curvature, 
generating intense bursts of gravitational radiation. These signals encode 
detailed information about the near-horizon geometry and the combined effects 
of QED corrections and dark matter.

The geodesics are solved in spherical coordinates $(r,\theta,\phi)$, but the 
waveforms are expressed in a Cartesian frame $(x,y,z)$ defined by 
\cite{Thorne:1980ru, Zhang:2025wni, BabakEtAl:2007}
\begin{equation}
x=r\sin\theta\cos\phi,\quad
y=r\sin\theta\sin\phi,\quad
z=r\cos\theta .
\end{equation}

Within the linearized approximation,
\begin{equation}
g_{\mu\nu}=\eta_{\mu\nu}+h_{\mu\nu},\qquad
\bar{h}^{\mu\nu}=h^{\mu\nu}-\frac{1}{2}h\,\eta^{\mu\nu},
\end{equation}
and imposing the Lorentz gauge $\bar{h}^{\mu\alpha}{}_{,\alpha}=0$, one obtains
\begin{equation}
\square \bar{h}^{\mu\nu}=-16\pi T^{\mu\nu}.
\end{equation}

In the slow-motion limit, the waveform reduces to the quadrupole formula,
\begin{equation}
\bar{h}^{ij}(t,\mathbf{x})=\frac{2}{D_L}
\left[\ddot{I}^{ij}(t-D_L)\right],
\end{equation}
with mass quadrupole moment \cite{Thorne:1980ru}
\begin{equation}
I^{ij}=\int x^i x^j T^{tt}(t,\mathbf{x})\, d^3x.
\end{equation}

For a point particle of mass $m$ following the trajectory $\mathbf{Z}(t)$,
\begin{equation}
T^{tt}(t,\mathbf{x})=
m\,\delta^3(\mathbf{x}-\mathbf{Z}(t)),
\end{equation}
leading to \cite{Zhao:2024exh,Meng:2024cnq}
\begin{equation}
h_{ij}=\frac{2m}{D_L}
\left(a_i x_j + a_j x_i + 2 v_i v_j \right),
\end{equation}
where $v_i$ and $a_i$ are the velocity and acceleration.

The detector frame $(X,Y,Z)$ is obtained by rotating the source frame through 
inclination $\iota$ and longitude of pericenter $\zeta$,
\begin{eqnarray}
\hat{e}_X &=& (\cos\zeta,-\sin\zeta,0),\\
\hat{e}_Y &=& (\sin\iota\sin\zeta,\cos\iota\cos\zeta,-\sin\iota),\\
\hat{e}_Z &=& (\sin\iota\sin\zeta,-\sin\iota\cos\zeta,\cos\iota),
\end{eqnarray}
and the two GW polarizations are \cite{Yang:2024lmj, Meng:2024cnq}
\begin{eqnarray}
h_+ &=& \frac{1}{2}(e_X^i e_X^j - e_Y^i e_Y^j) h_{ij},\\
h_\times &=& \frac{1}{2}(e_X^i e_Y^j - e_Y^i e_X^j) h_{ij}.
\end{eqnarray}

We consider a representative EMRI with a compact object of mass 
$m=10M_\odot$ orbiting a supermassive EH--PFDM black hole of mass 
$M=10^6M_\odot$, inclination $\iota=\pi/4$, $\zeta=\pi/4$, and luminosity 
distance $D_L=200\,{\rm Mpc}$.

\begin{figure*}[!htbp]
\centering
\includegraphics[scale=0.45]{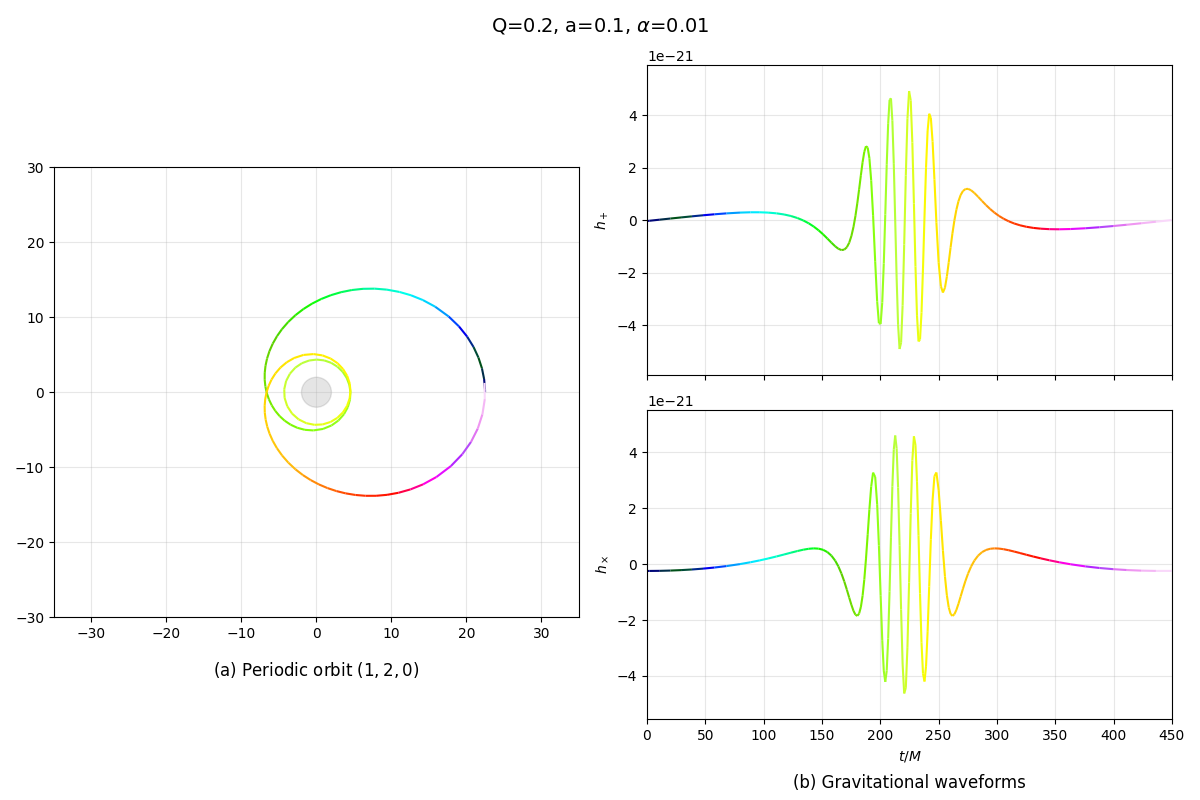}
\caption{Gravitational waveform associated with the periodic orbit 
$(z,w,\nu)=(1,2,0)$ in the EH--PFDM spacetime with $Q=0.2$, $a=0.1$, 
$\alpha=0.01$. The left panel shows the corresponding periodic orbit, 
while the right panels display the GW polarizations $h_+$ and $h_\times$.}
\label{fig:gw1}
\end{figure*}

\begin{figure*}[!htbp]
\centering
\includegraphics[scale=0.45]{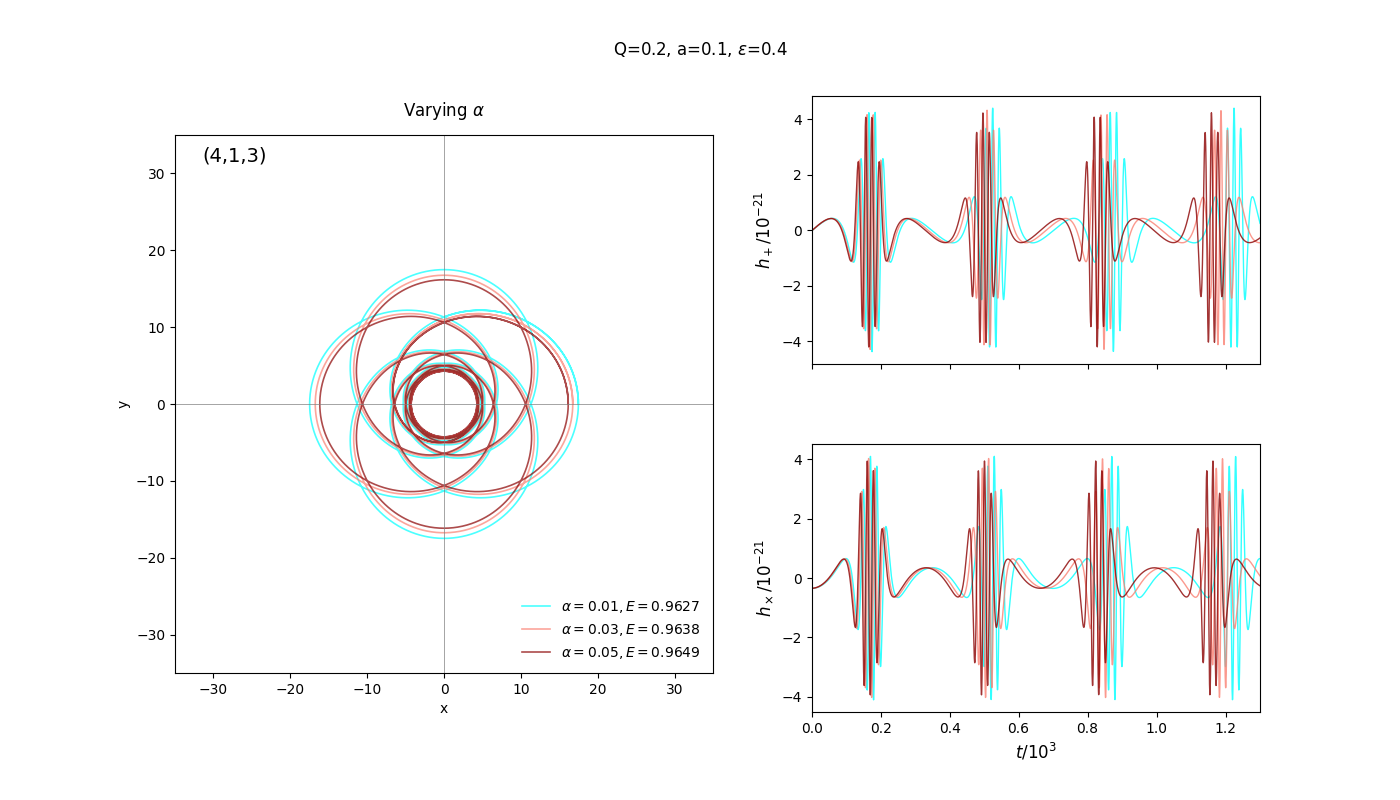}
\caption{Effect of varying the dark matter parameter $\alpha$ for 
$(z,w,\nu)=(4,1,3)$ with $Q=0.2$, $a=0.1$, and $\epsilon=0.4$. Increasing 
$\alpha$ deforms the orbital geometry and suppresses the GW amplitude.}
\label{fig:gw2}
\end{figure*}

\begin{figure*}[!htbp]
\centering
\includegraphics[scale=0.45]{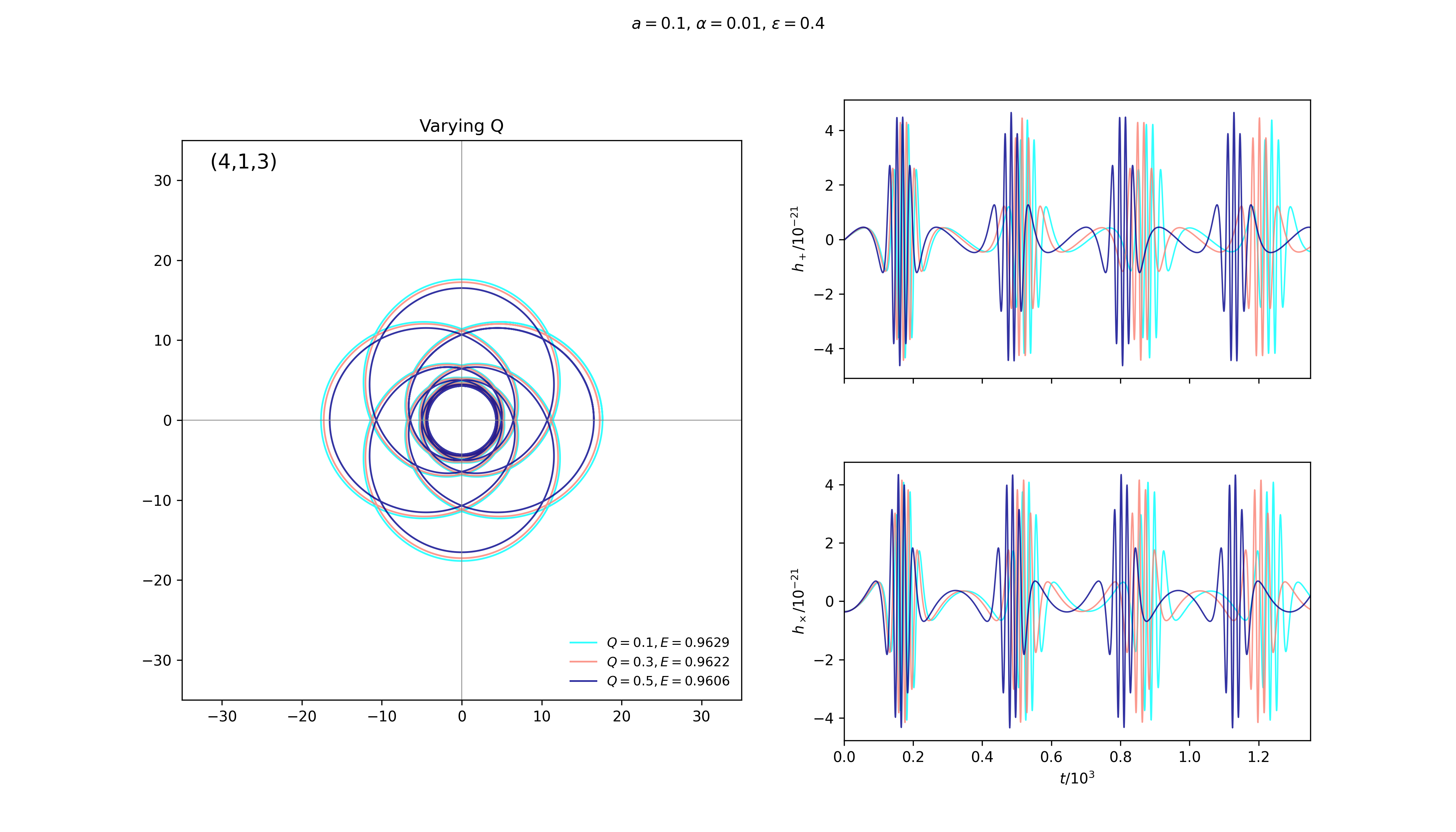}
\caption{Effect of varying the electric charge $Q$ on periodic orbits 
and waveforms in the EH--PFDM spacetime with fixed $a=0.1$ and 
$\alpha=0.01$. Larger $Q$ enhances whirl activity and amplifies 
high-frequency GW components.}
\label{fig:gw3}
\end{figure*}

\begin{figure*}[!htbp]
\centering
\includegraphics[scale=0.45]{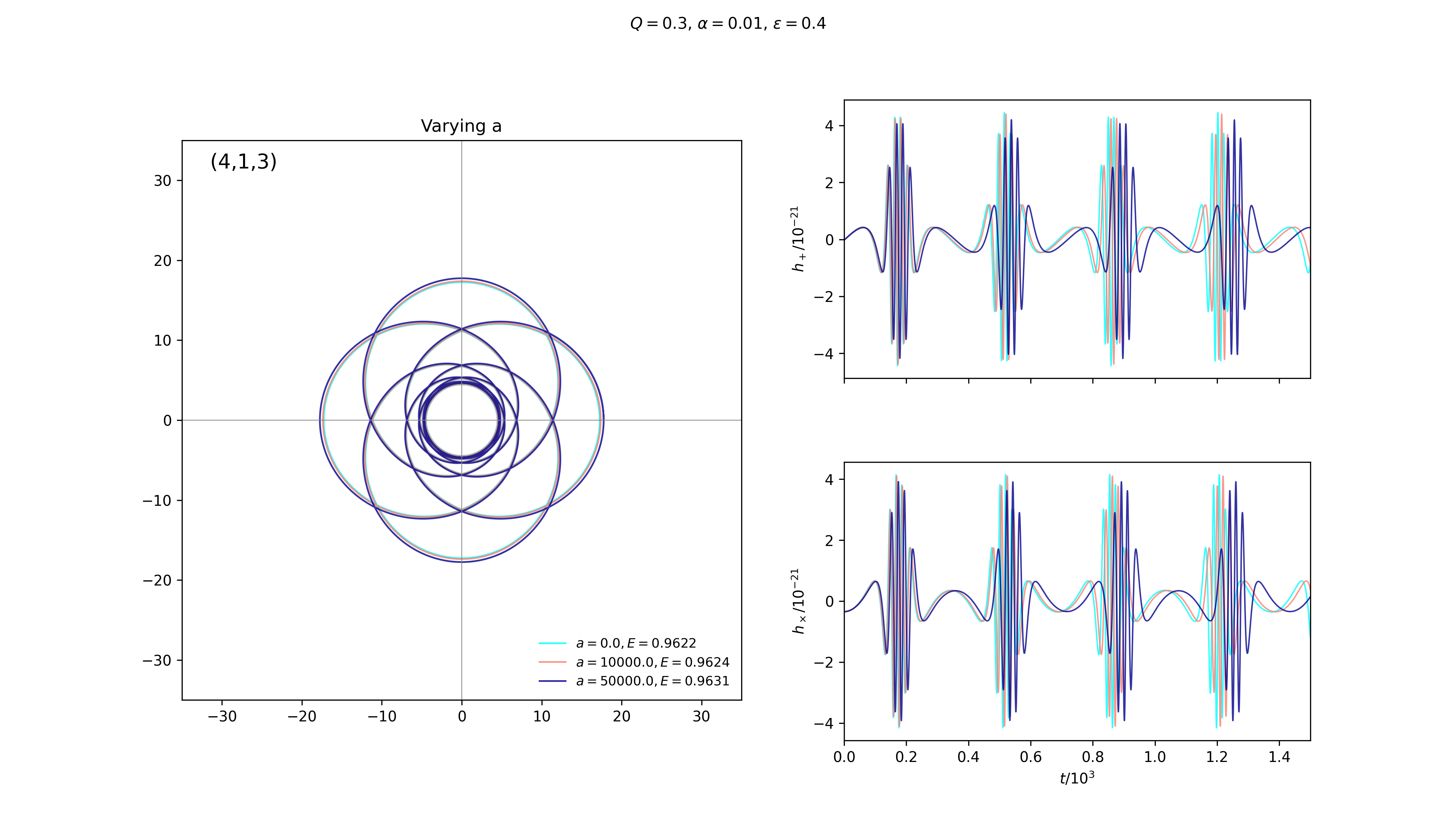}
\caption{Effect of varying the QED parameter $a$ for fixed 
$Q=0.3$, $\alpha=0.01$, and $\epsilon=0.4$. The QED correction mainly 
modifies the near-horizon motion, leading to moderate but systematic 
changes in the waveform amplitude and frequency.}
\label{fig:gw4}
\end{figure*}

The waveforms in Figs.~\ref{fig:gw1}--\ref{fig:gw4} exhibit a clear imprint 
of the underlying zoom--whirl orbital structure. The smooth, low-amplitude 
segments correspond to zoom phases, while the rapidly oscillating, 
high-amplitude bursts arise from whirl phases near the unstable circular 
orbit.

Figure~\ref{fig:gw1} illustrates a representative gravitational waveform 
associated with the periodic orbit $(z,w,\nu)=(1,2,0)$ in the EH--PFDM 
spacetime. The left panel shows that the trajectory consists of a single 
extended zoom segment followed by two tight whirls near the black hole. 
During the zoom phase, the particle follows a highly elongated orbit in the 
weak-field region, and the corresponding gravitational radiation remains 
smooth and of low amplitude. In contrast, the whirl phase is characterized 
by rapid revolutions near the unstable circular orbit, where the particle 
experiences strong acceleration in the high-curvature region. This produces 
the sharp, high-frequency oscillations observed in both GW polarizations 
$h_{+}$ and $h_{\times}$. The clustering of large-amplitude peaks in the 
waveform coincides precisely with the time interval during which the 
particle is temporarily trapped near the black hole.

Figure~\ref{fig:gw2} shows the effect of varying the dark matter parameter 
$\alpha$ for a fixed topological class $(z,w,\nu)=(4,1,3)$. As $\alpha$ 
increases, the overall size of the orbit grows and the periapsis shifts 
outward, reflecting the effective weakening of the gravitational potential 
induced by dark matter. This geometric deformation leads to a systematic 
suppression of the gravitational wave amplitude in both polarizations. 
Physically, the particle spends less time in the strong-field region, and 
its acceleration during the whirl phases is reduced, resulting in weaker 
radiation. Although the qualitative zoom--whirl structure of the signal is 
preserved, the high-frequency components become less pronounced as 
$\alpha$ increases.

The influence of the electric charge $Q$ is illustrated in 
Fig.~\ref{fig:gw3}. For fixed $\alpha$ and $a$, increasing $Q$ enhances the 
whirl activity and leads to tighter inner loops in the orbital geometry. 
This produces a clear amplification of the high-frequency components in 
the gravitational waveform. The particle experiences stronger curvature 
effects near the unstable circular orbit, leading to more rapid oscillations 
and larger peak amplitudes during the whirl phases. As a result, the 
electric charge acts as an efficient amplifier of strong-field 
gravitational radiation.

Figure~\ref{fig:gw4} displays the effect of varying the QED parameter $a$ 
while keeping $Q$ and $\alpha$ fixed. In this case, the modifications of 
the orbital geometry are more subtle than those produced by varying 
$\alpha$ or $Q$, indicating that QED corrections mainly affect the 
near-horizon region without significantly altering the global shape of 
the trajectory. Nevertheless, the waveforms exhibit systematic shifts in 
both amplitude and frequency, with larger values of $a$ producing slightly 
stronger and more rapidly oscillating signals. This behavior reflects the 
sensitivity of near-horizon motion to quantum electrodynamic corrections, 
which remain imprinted in the detailed temporal structure of the 
gravitational radiation.

\section{Conclusion}\label{sec05}

In this paper, we studied equatorial periodic timelike orbits and their associated gravitational wave signals in the spacetime of an Euler--Heisenberg black hole surrounded by perfect fluid dark matter. The analysis was designed to isolate the combined role of two distinct sources of deviation from the standard charged black hole picture: the short-distance modification induced by Euler--Heisenberg nonlinear electrodynamics and the environmental deformation produced by the dark-matter sector. To this end, we examined the geodesic structure of the EH--PFDM geometry, determined the principal orbital thresholds relevant for bound motion, classified periodic trajectories through the rational parameter and the $(z,w,\nu)$ taxonomy, and constructed representative waveforms using the numerical kludge approach.

Our results show that the EH--PFDM background supports a rich spectrum of bound periodic motion in the strong-field regime, including pronounced zoom--whirl trajectories. The structure of the effective potential is modified in a systematic way by the model parameters, and these modifications are reflected in the locations of the marginally bound orbit and the innermost stable circular orbit. In particular, the perfect fluid dark matter parameter changes the global shape of the effective potential and shifts the characteristic orbital radii, thereby altering the domain of admissible bound motion. By contrast, the Euler--Heisenberg correction acts more strongly in the inner region of the spacetime and mainly affects the portion of the motion that probes the near-horizon geometry.

The periodic-orbit analysis further shows that the topological structure of relativistic bound motion remains a sensitive diagnostic of the underlying spacetime. The rational parameter captures the transition from moderately precessing bound motion to the near-critical zoom--whirl regime, while the $(z,w,\nu)$ classification provides a transparent description of how orbital morphology changes as the conserved quantities and model parameters are varied. In this sense, periodic orbits serve not only as a convenient organizational tool for relativistic dynamics, but also as a bridge between the geometry of the spacetime and potentially observable features of gravitational radiation.

The waveform analysis indicates that the imprint of the orbital structure survives clearly in the corresponding numerical-kludge signals. The zoom segments generate relatively smooth and weakly modulated radiation, whereas the whirl episodes near the unstable circular orbit produce the burst-like, rapidly oscillating features that dominate the strong-field part of the signal. Within the framework considered here, the dark-matter parameter tends to move the motion outward and reduce the overall waveform amplitude, while the electric charge and the Euler--Heisenberg correction primarily influence the strong-field portion of the orbit and the associated high-frequency structure of the radiation. The resulting waveform morphology therefore reflects, in a direct way, the interplay between environmental effects and nonlinear electrodynamic corrections.

The present analysis should be viewed as a first step toward understanding EMRI-like dynamics in black hole spacetimes influenced simultaneously by nonlinear electrodynamics and dark matter. Since the waveforms were constructed in the numerical kludge approximation and radiation reaction was neglected over each orbital cycle, the results are best interpreted as capturing qualitative and semi-quantitative trends rather than providing precision templates for data analysis. Nevertheless, they demonstrate that periodic orbits in the EH--PFDM spacetime offer a useful theoretical laboratory for identifying how strong-field geodesic structure is encoded in gravitational wave signals.

There are several natural directions for future work. A more complete treatment should incorporate dissipative evolution and self-force effects in order to follow inspiral trajectories beyond the periodic-orbit approximation. It would also be interesting to extend the analysis to rotating generalizations of the background, where frame dragging would enrich the orbital taxonomy and the waveform structure. On the observational side, a systematic comparison between shadow, lensing, QPO, and ringdown constraints on the same class of spacetimes may help clarify which strong-field channels are most sensitive to environmental matter and nonlinear electromagnetic corrections. We hope that the present results will be useful for such future investigations of black hole dynamics beyond the simplest vacuum scenarios.

\section*{Acknowledgments}
DJG acknowledges the contribution of the COST Action CA21136  -- ``Addressing observational tensions in cosmology with systematics and fundamental physics (CosmoVerse)".  A. \"O. would like to acknowledge networking support of the COST Action CA21106 - COSMIC WISPers in the Dark Universe: Theory, astrophysics and experiments (CosmicWISPers), the COST Action CA22113 - Fundamental challenges in theoretical physics (THEORY-CHALLENGES), the COST Action CA21136 - Addressing observational tensions in cosmology with systematics and fundamental physics (CosmoVerse), the COST Action CA23130 - Bridging high and low energies in search of quantum gravity (BridgeQG), and the COST Action CA23115 - Relativistic Quantum Information (RQI) funded by COST (European Cooperation in Science and Technology). A. \"O. also thanks to EMU, TUBITAK, ULAKBIM (Turkiye) and SCOAP3 (Switzerland) for their support.

\section*{Data Availability Statement}
There are no new data associated with this article.

\bibliography{ref}

\end{document}